\documentclass{article}

\usepackage[authoryear, compress]{natbib}

\usepackage{neurips_2024} 

\usepackage[utf8]{inputenc} 
\usepackage[T1]{fontenc}    
\usepackage{hyperref}       
\usepackage{url}            
\usepackage{booktabs}       
\usepackage{amsfonts}       
\usepackage{nicefrac}       
\usepackage{microtype}      
\usepackage{xcolor}         
\usepackage{tikz}
\usetikzlibrary{positioning, shapes.geometric, decorations.markings, arrows.meta}

\tikzset{
  midarrow/.style args={#1}{
    decoration={
      markings,
      mark=at position #1 with {\arrow{Stealth}}
    },
    postaction={decorate}
  }
}

\hypersetup{
    colorlinks=true,    
    citecolor=blue,     
    linkcolor=blue,     
    urlcolor=blue       
}

\usepackage{amsmath,amssymb} 
\usepackage{algorithmic}
\usepackage{graphicx}
\usepackage{textcomp}
\usepackage{listings}

\newcommand{\code}[1]{\texttt{#1}}

\title{LeanExplore: A search engine for Lean 4 declarations}

\author{%
  Justin Asher \\
  Independent Researcher \\
  \texttt{justinchadwickasher@gmail.com} \\
}

\begin{document}

\maketitle

\begin{abstract}
The expanding Lean 4 ecosystem poses challenges for navigating its vast libraries. This paper introduces LeanExplore, a search engine for Lean 4 declarations. LeanExplore enables users to semantically search for statements, both formally and informally, across select Lean 4 packages\footnote{Currently indexed packages include: Batteries, Init, Lean, Mathlib, PhysLean, and Std.}. This search capability is powered by a hybrid ranking strategy, integrating scores from a multi-source semantic embedding model (capturing conceptual meaning from formal Lean code, docstrings, AI-generated informal translations, and declaration titles), BM25+ for keyword-based lexical relevance, and a PageRank-based score reflecting declaration importance and interconnectedness. The search engine is accessible via a dedicated website\footnote{\url{https://www.leanexplore.com/}} and a Python API\footnote{\url{https://github.com/justincasher/lean-explore}}. Furthermore, the database can be downloaded, allowing users to self-host the service. LeanExplore integrates easily with LLMs via the model context protocol (MCP), enabling users to chat with an AI assistant about Lean declarations or utilize the search engine for building theorem-proving agents. This work details LeanExplore's architecture, data processing, functionalities, and its potential to enhance Lean 4 workflows and AI-driven mathematical research.
\end{abstract}

\section{Introduction}

Lean 4 \citep{Lean4} is a type-theoretic programming language and interactive theorem prover, enabling users to formally verify the correctness of mathematical theorems and complex software systems. Its ecosystem has seen significant expansion, largely driven by comprehensive libraries such as Mathlib \citep{Mathlib}, which formalizes a vast corpus of graduate-level mathematics, and emerging specialized packages like PhysLean for physics \citep{PhysLean}. This growth, while a testament to Lean's power and community engagement, introduces substantial challenges in navigating the extensive collection of existing code and formalized knowledge.

The task of formalizing new mathematical statements or developing verified software often requires users to locate, understand, and utilize numerous existing definitions, theorems, and tactics—a discovery process that can be particularly arduous, even for experienced researchers. Effectively navigating these large codebases is frequently hampered because not all declarations are accompanied by descriptive docstrings, and conventional keyword-based search methods often prove inadequate. For instance, a naive search for a well-known concept like the "fundamental theorem of calculus" within the extensive Mathlib documentation \citep{MathlibDocs} yields no direct results. This example illustrates the significant difficulty in bridging the gap between conceptual mathematical ideas and their specific, formal representations in Lean, especially when attempting to search using natural language.

To mitigate these difficulties, we introduce LeanExplore, a search engine for Lean 4 declarations. Implemented both as a website and a Python package, LeanExplore offers robust semantic search capabilities alongside a suite of specialized tools. These resources are engineered to significantly enhance how users locate, understand, and interact with the vast and growing body of Lean code. Crucially, the system design facilitates not only more intuitive human engagement with Lean but also the effective integration of its formal capabilities as a powerful tool for artificial intelligence.

This paper makes several key contributions:
\begin{itemize}
    \item We introduce a hybrid search methodology tailored for the Lean 4 domain, significantly improving declaration retrieval. At its core, this approach ranks declarations by combining scores from three distinct signals: multi-source semantic embeddings (derived from formal Lean code, docstrings, AI-generated informal translations, and declaration titles), BM25+ for keyword-based lexical relevance, and a PageRank-based score reflecting declaration importance and interconnectedness. This integrated ranking strategy enables users to interact with the database semantically, through keywords, or by known titles, offering versatile and effective discovery.
        
    \item We detail a flexible system architecture that supports two distinct operational modes: local data access for offline use and maximal user control, and a remote API mode for zero-setup access to centrally managed datasets.

    \item We describe the design and implementation of a robust data pipeline that transforms Lean 4 source code into a structured and semantically rich knowledge base. This pipeline introduces \code{StatementGroup}s, a novel abstraction for grouping related declarations. Furthermore, it employs an LLM-driven, inductive process that leverages inter-declaration dependencies to generate high-quality informal English translations for formal statements, significantly enhancing the data for effective semantic search.
    
    \item We introduce a model context protocol (MCP) server implementation, which exposes LeanExplore's functionalities as a set of tools consumable by external AI agent systems. This feature aims to foster new avenues for AI research in formal mathematics by allowing AI to directly leverage Lean's structured knowledge. We demonstrate the effectiveness of this MCP through an AI chatbot that users can access via the CLI.
\end{itemize}
The remainder of this paper is structured as follows: Section \ref{sec:related_work} provides an overview of relevant background and related work. Section \ref{sec:architecture} elaborates on the system architecture, focusing on its data processing and search capabilities. Section \ref{sec:features} describes the core features and functionalities. Section \ref{sec:use_cases} discusses various use cases and illustrative scenarios. Section \ref{sec:experimental_results} presents the experimental evaluation of LeanExplore's search efficacy. Finally, Section \ref{sec:future_work} outlines future work and directions, and Section \ref{sec:conclusion} concludes the paper.

\section{Background and related work}
\label{sec:related_work}

The ability to search formal mathematics databases is not a unique challenge for Lean. Users of many other interactive theorem provers (ITPs), such as Coq, Isabelle, and Agda, have also faced this navigation difficulty \citep{Winkler2014FocusGroups}. While many ITPs offer built-in search commands, these often require users to guess specific patterns or names and may be limited to currently loaded theories. To address this, external tools have emerged. For instance, Isabelle’s Sledgehammer connects to external automated theorem provers (ATPs) to find relevant lemmas and proofs \citep{Blanchette2011SledgehammerDisproof}, and its FindFacts tool indexes the entire corpus for text-based search \citep{Hou2022FindFacts}. Agda’s ``Search About'' command leverages its rich type system for contextual search \citep{AgdaManualSearchAbout}. Within the Lean \citep{Lean4} ecosystem itself, prior to LeanExplore, tools like Loogle provided pattern-based search for Mathlib \citep{Mathlib, LoogleWebsite}, and more recent semantic search engines like Moogle.ai \citep{MoogleAI} and LeanSearch \citep{Gao2024MathlibSearch} have applied embeddings to allow natural language queries over Mathlib.

The evolution of code search provides important parallels. Early systems often treated code as plain text, allowing keyword or regular expression queries. However, the structured nature of code led to the development of syntax-aware and semantic search techniques. Tools like Hoogle for Haskell, which enables searching by type signature, demonstrated the power of leveraging language-specific semantics \citep{Mitchell2008Hoogle}. Searching formal languages presents unique challenges due to their precise syntax and semantics, where minor textual variations can drastically alter meaning, necessitating high precision and an understanding of logical structure.

Semantic search, which aims to understand user intent and contextual meaning rather than just matching keywords, has become pivotal. This is typically powered by embeddings---dense vector representations where semantic similarity corresponds to proximity in the vector space \citep{EmbeddingsReview}. While general text embeddings (e.g., from sentence transformers \citep{Reimers2019SentenceBERT}) provide a baseline, there’s a strong trend towards specialized embeddings for code (e.g., CodeBERT \citep{Feng2020CodeBERT}) and formal mathematical content. For instance, \citet{Tao2025PremiseRetriever} developed a BERT-based model specifically trained to embed Lean proof states and premises, and various works explore contrastive learning to train embeddings for mathematical formulas and structures without extensive labeled data \citep{Wu2023CLFE}. LeanExplore’s multi-source semantic embedding strategy aligns with this trend, seeking to capture the semantics of both formal Lean statements and informal descriptions.

The intersection of artificial intelligence and formal mathematics is rapidly accelerating, driven by LLMs which now routinely assist in and increasingly automate tasks like automated theorem proving (ATP). ATP examples include modular construction (LEGO-Prover \citep{Wang2023LegoProver}), hybrid methods (HybridProver \citep{Aglot2025HybridProver}), RL-driven whole-proof generation (Kimina-Prover \citep{kimina_prover_2025}), type inhabitation solvers (Canonical \citep{Norman2025Canonical}), and specialized neuro-symbolic systems achieving expert performance in areas like Olympiad geometry (AlphaGeometry2 \citep{Chervonyi2025AlphaGeometry2}). LLMs also enhance tactic prediction and premise selection in interactive theorem provers (ITPs), leveraging platforms like LeanDojo with its ReProver model \citep{LeanDojo} and interfaces such as Pantograph \citep{pantograph}, and enable full proof synthesis, for instance with "draft, sketch, and prove" (DSP) \citep{Jiang2023DraftSketchProve}. This maturation includes LLMs bridging natural language and formal systems via autoformalization, with examples including initial efforts \citep{Wu2022Autoformalization} and projects like the Lean Workbook \citep{Alonzo2024LeanWorkbook} and the FormalMATH pipeline \citep{Azerbayev2025FormalMATH}. Mirroring AI-powered developer copilots (e.g., GitHub Copilot) in programming, similar capabilities are emerging in ITPs (Lean Copilot \citep{Han2024LeanCopilot}; LLMSTEP \citep{Achan2025LLMSTEP}; PALM for Coq \citep{Minghai2024PALM}). This adaptation is logical, as formal proofs can be seen as specialized code \citep{LeanDojo}, signifying a paradigm shift in advancing formal mathematics.

Anthropic's recently open-sourced model context protocol (MCP) \citep{anthropic_model_context_protocol} aims to standardize connections between AI assistants and tools. For the Lean theorem prover, MCP implementations are emerging. For instance, "lean-lsp-mcp" \citep{lean_lsp_mcp} facilitates agentic interaction by exposing Lean's language server protocol features and project analysis tools (such as diagnostics, goal states, and hover information). Similarly, "LeanTool" \citep{lean_tool} is designed to let LLMs interact with Lean as a "Code Interpreter" for syntax checking and interactive feedback. LeanExplore builds on these tools using the MCP to allow models to query for Lean declarations.

\section{LeanExplore system architecture}
\label{sec:architecture}

The architecture of LeanExplore is designed to be lightweight and modular. It allows users to easily query and interact with the database locally and through a dedicated API.

\subsection{Lean data extraction}
\label{sec:lean_data_extraction}

LeanExplore employs a two-pronged approach for extracting raw data from Lean 4 libraries. First, a global analysis of compiled target library environments gathers comprehensive metadata for all formal declarations (e.g., theorems, definitions). This includes their properties, types, and direct inter-dependencies across the entire codebase, yielding structured data files that form an initial project overview.

Complementing this global view, a second process, adapted from LeanDojo \citep{LeanDojo}, performs a detailed analysis of individual source files. This captures fine-grained information such as abstract syntax trees (ASTs) for each command, tactic application traces with pre- and post-goal states, and premise usage within proofs. This fine-grained data, particularly the ASTs (essential for locating specific source text fragments) and the detailed inter-declaration dependencies, is crucial for subsequent processing steps. While some detailed trace data (like tactic states) is preserved for potential future features, such as proof state search \citep{Tao2025PremiseRetriever}, the immediate focus for LeanExplore is on the declarations themselves and their relationships.

\subsection{StatementGroups: Organizing declarations}
\label{sec:statement_groups}

To present information in a manner that aligns with how users author and perceive Lean code, we introduce the \code{StatementGroup} abstraction. This is necessary because single high-level Lean constructs (e.g., \code{structure}, \code{inductive} type definitions, or even a single \code{theorem} command) often elaborate into multiple, lower-level underlying declarations in the compiled Lean environment. Treating each of these generated items independently in a search interface would fragment results and misrepresent the structure of the user-written code.

Each \code{StatementGroup} thus represents a unique, contiguous block of source code as authored by the user. All individual declarations originating from such a block are associated with their respective \code{StatementGroup}. For each group, one member is designated as the primary declaration, and its Lean name and docstring (if present) serve as the canonical name and docstring for the entire group.

For example, a single user-authored block in Lean, such as the following \texttt{structure} definition for a scheme in algebraic geometry:
\begin{quote}
\centering
$\begin{array}{@{}l@{}}
\texttt{structure Scheme extends LocallyRingedSpace where} \\
\quad \texttt{local\_affine :} \\
\quad\quad \forall x : \texttt{toLocallyRingedSpace,} \\
\quad\quad\quad \exists (U : \texttt{OpenNhds } x) (R : \texttt{CommRingCat}), \\
\quad\quad\quad\quad \texttt{Nonempty} \\
\quad\quad\quad\quad\quad (\texttt{toLocallyRingedSpace.restrict } U\texttt{.isOpenEmbedding} \cong \\
\quad\quad\quad\quad\quad\quad \texttt{Spec.toLocallyRingedSpace.obj (op } R\texttt{)})
\end{array}$
\end{quote}
This piece of code actually elaborates into multiple distinct declarations within the Lean environment. These include, among others:
\begin{itemize}
    \item \texttt{AlgebraicGeometry.Scheme} (an \texttt{inductive} type)
    \item \texttt{AlgebraicGeometry.Scheme.mk} (its \texttt{constructor})
    \item \texttt{AlgebraicGeometry.Scheme.local\_affine} (the \texttt{theorem} for the local affine condition)
    \item \texttt{AlgebraicGeometry.Scheme.toLocallyRingedSpace} (a \texttt{definition} for the projection)
    \item \texttt{AlgebraicGeometry.Scheme.rec} (its \texttt{recursor})
    \item \texttt{AlgebraicGeometry.Scheme.noConfusion} (an auxiliary \texttt{definition})
\end{itemize}
The \code{StatementGroup} abstraction correctly groups all these generated declarations (along with others like \texttt{casesOn} and \texttt{recOn}) under the original user-written \texttt{structure Scheme} block. This approach facilitates a more user-familiar presentation of search results, provides coherent units for subsequent analysis, and still preserves access to the fine-grained details of each underlying declaration.

\subsection{Dependency analysis and data enrichment}
\label{sec:dependency_analysis}

The fine-grained inter-declaration dependencies extracted during the initial data processing (as described in Section~\ref{sec:lean_data_extraction}) are aggregated to construct a directed dependency graph at the \code{StatementGroup} level. In this graph, an edge exists from \code{StatementGroup} A to \code{StatementGroup} B if any declaration within A depends on any declaration within B. This \code{StatementGroup}-level dependency graph is pivotal for two subsequent data enrichment processes that significantly enhance searchability and relevance ranking.

First, the graph enables an inductive approach to generating informal English counterparts for each \code{StatementGroup}, which are essential for effective semantic search. We traverse the dependency graph in a topological order (processing dependencies before their dependents). For each \code{StatementGroup}, Google's Gemini 2.0 Flash model is prompted with its formal Lean code, its docstring (if available), and, crucially, the already-generated informal English translations of its direct dependencies (i.e., the \code{StatementGroup}s it points to in the graph). This contextual information allows the LLM to produce more accurate and consistent natural language descriptions by understanding the statement in the context of related, already-translated concepts.

Second, this same \code{StatementGroup}-level dependency graph also serves as the basis for assessing the relative importance and interconnectedness of statements within the indexed libraries. We compute log-transformed PageRank scores \citep{page1998pagerank} for each \code{StatementGroup} using this graph. These PageRank scores provide a quantitative measure of a statement's centrality and influence, subsequently serving as a key signal in LeanExplore's search ranking algorithm (detailed in Section~\ref{sec:search_algorithm}) to help prioritize more authoritative or foundational results.

\subsection{Search algorithm}
\label{sec:search_algorithm}

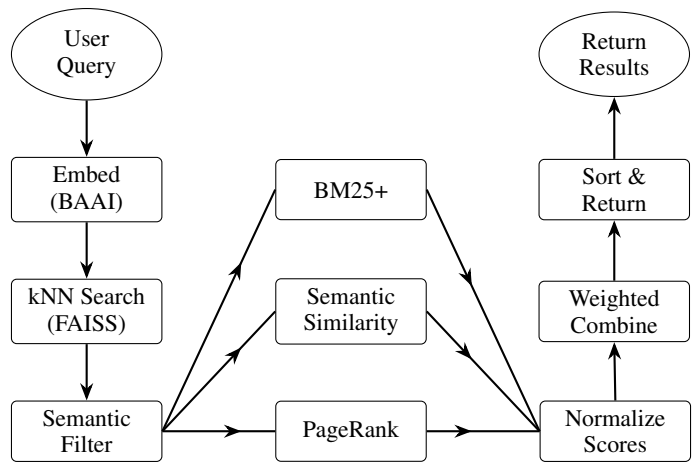
\begin{figure}[!b]
  \centering
  \begin{tikzpicture}[
      process/.style={
        rectangle, draw, rounded corners=2pt,
        minimum width=2cm, minimum height=0.8cm,
        align=center, font=\footnotesize
      },
      io/.style={
        ellipse, draw,
        minimum width=2cm, minimum height=0.8cm,
        align=center, font=\footnotesize
      },
      arrow/.style={-{Stealth}, thick},
      dashedarrow/.style={-{Stealth}, thick, dashed}
    ]
    \node[io]                            (query)   {User\\Query};
    \node[process,below=0.75cm of query]   (embed)   {Embed\\(BAAI)};
    \node[process,below=0.75cm of embed]   (knn)     {kNN Search\\(FAISS)};
    \node[process,below=0.75cm of knn]     (filter)  {Semantic\\Filter};

    \node[process, right=1.5cm of embed] (bm25) {BM25+};
    \node[process, right=1.5cm of knn] (sem)  {Semantic\\Similarity};
    \node[process, right=1.5cm of filter]   (pr)   {PageRank};

    \node[process, right=1.5cm of pr]            (norm)    {Normalize\\Scores};

    \node[process, right=5cm of knn]            (combine) {Weighted\\Combine};
    \node[process, right=5cm of embed]         (sort)    {Sort \&\\Return};
    \node[io, right=5cm of query]                  (output) {Return\\Results};

    \coordinate (merge) at (norm.west);

    \draw[arrow] (query)  -- (embed);
    \draw[arrow] (embed)  -- (knn);
    \draw[arrow] (knn)    -- (filter);

    \draw[-, thick, midarrow=0.7]  (filter.east) -- (bm25.west);
    \draw[-, thick, midarrow=0.7]  (filter.east) -- (sem.west);
    \draw[-, thick, midarrow=0.7]  (filter.east) -- (pr.west);

    \draw[-, thick, midarrow=0.4] (bm25.east) to[right=20] (merge);
    \draw[-, thick, midarrow=0.4] (sem.east)  -- (merge);
    \draw[-, thick, midarrow=0.4] (pr.east)   to[left=20] (merge);


    \draw[arrow] (norm.north)      -- (combine.south);
    \draw[arrow] (combine.north)   -- (sort.south);
    \draw[arrow] (sort.north)      -- (output.south);
  \end{tikzpicture}
  \caption{LeanExplore hybrid search pipeline.}
  \label{fig:leanexplore_pipeline}
\end{figure}

LeanExplore employs a hybrid search algorithm, synergizing semantic understanding, lexical matching (BM25+), and structural importance (PageRank) to retrieve and rank \code{StatementGroup}s. This multi-stage process holistically assesses relevance.

Each \code{StatementGroup} is represented by multiple embeddings corresponding to distinct textual facets: its name, its docstring (if available), and a composite "informal statement." This informal statement is constructed by concatenating an AI-generated natural language translation of the formal content with contextual keywords derived from the \code{StatementGroup}'s file path (e.g., the path \code{Mathlib/FieldTheory/PolynomialGaloisGroup.lean} contributes keywords "field theory polynomial galois group"). All these textual components are embedded using BAAI's bge-base-en-v1.5 model \citep{BGEManual}. For efficient k-nearest neighbors (kNN) retrieval, these embeddings are indexed using FAISS \citep{FAISS}, specifically an inverted file structure with 4096 quantization cells (clusters). 

User queries are embedded using the same BAAI model. The kNN search first identifies relevant cells and then performs exact distance calculations only for vectors within those cells. Raw FAISS distances are converted to a [0,1] similarity scale, and candidates are filtered by a configurable semantic similarity threshold (default: 0.525).

For \code{StatementGroup}s that pass this initial semantic filter, two additional relevance signals are computed. Firstly, a lexical relevance score is determined using the BM25+ algorithm \citep{lv2011lower}. This score measures term-based similarity between the user's query and a specially constructed textual representation for each candidate \code{StatementGroup}, which amalgamates its primary declaration's name, docstring, informal descriptions, the formal display statement text, and a processed version of its source file path. Both query and candidate texts undergo standard tokenization (lowercasing) before BM25+ scoring. Secondly, each candidate \code{StatementGroup} contributes its pre-calculated PageRank score, derived from the inter-dependencies among \code{StatementGroup}s, reflecting its structural importance within the broader codebase.

To combine these diverse signals, the semantic similarity scores, BM25+ lexical scores, and PageRank scores for the current set of filtered candidates are each independently min-max normalized to a [0,1] range. This local normalization ensures equitable contribution from each signal. The final relevance score for each \code{StatementGroup} is then calculated as a weighted linear combination of these three normalized scores, with configurable weights (defaults: semantic similarity 1.0, BM25+ 1.0, PageRank 0.2). A detailed example illustrating this scoring process, including a comparison of results with and without the BM25+ component, is provided in Appendix~\ref{appendix:scoring_example_comparison}. Finally, the \code{StatementGroup}s are presented to the user, sorted in descending order by this composite score.

\section{Core features and functionality}
\label{sec:features}
LeanExplore offers a suite of features designed to enhance the discovery, understanding, and utilization of Lean 4 code, accessible through multiple interfaces.

\subsection{Web interface}
\label{subsec:web_interface}
The primary public interface to LeanExplore is its dedicated website.\footnote{\url{https://www.leanexplore.com}}. It provides:
\begin{itemize}
    \item \textbf{Semantic and direct search:} Users can perform semantic searches using natural language (e.g., "find theorems about continuous functions") or Lean-like syntax to retrieve relevant \code{StatementGroup}s. The interface also supports direct lookup of declarations by their fully qualified Lean names and retrieval of items by their unique system IDs. Search results can be filtered by specific Lean packages (e.g., Mathlib, Batteries).
    \item \textbf{Dependency exploration:} For any retrieved \code{StatementGroup}, users can explore its direct dependencies and dependents, aiding in understanding code context and navigating project structures.
    \item \textbf{User accounts and API key management:} Registered users can view their search history. The website also serves as the portal for obtaining API keys, which enable access to LeanExplore's capabilities via its remote API for programmatic use (see Section \ref{subsec:python_api}).
    \item \textbf{Documentation:} Comprehensive documentation for LeanExplore, including API references and usage guides, is hosted on the website.
\end{itemize}

\subsection{Python library}
\label{subsec:python_api}
LeanExplore is also distributed as a Python package, providing both a command-line interface (CLI) and a library for programmatic integration. This package supports two operational modes:
\begin{itemize}
    \item \textbf{Local mode:} Users can download the pre-processed data for indexed Lean libraries and run LeanExplore entirely locally. This offers maximum control and offline access. The \code{Service} class provides synchronous access to search, ID retrieval, and dependency exploration functionalities against the local data.
    \item \textbf{Remote API mode:} Alternatively, users can leverage the centrally hosted LeanExplore service via an API key (obtained from the website). The \code{Client} offers asynchronous access to the same functionalities, requiring minimal local setup.
\end{itemize}
Both interfaces are designed to return data in consistent Pydantic model structures, simplifying client-side development.

\subsection{CLI}
\label{subsec:python_cli}
The \code{leanexplore} command-line interface (CLI) serves as the main entry point for managing local data, performing searches, getting \code{StatementGroup}s by their id, analyzing dependencies, and interacting with other system components.

The CLI enables users to interact with AI agents. The \code{leanexplore chat} CLI command launches an interactive AI-assisted session. This feature allows users to converse with an AI assistant that has access to the LeanExplore data backend (either local or remote via API key). This facilitates a more intuitive way to learn about Lean declarations, understand code, and explore the formal libraries.

Furthermore, the \code{leanexplore} CLI allows users to start a model context protocol (MCP) server (via \code{mcp server start}). This exposes LeanExplore's search and retrieval functionalities as tools consumable by external AI agent systems, enabling programmatic interaction for tasks like automated theorem proving or AI-driven mathematical research.

\section{Use cases and illustrative scenarios}
\label{sec:use_cases}
LeanExplore is designed to cater to diverse users and tasks within the Lean 4 ecosystem, enhancing productivity and enabling new research avenues.

\subsection{Accelerating formalization}
Consider a mathematician working on a new formal proof in Lean. They might recall a necessary lemma or concept but struggle to find its precise name or formulation within the vast Mathlib library. Conventional keyword searches might fail if the naming conventions are unknown or if no docstring perfectly matches their informal understanding. With LeanExplore, the mathematician can use a natural language query describing the desired mathematical property (e.g., "algebraically closed fields are infinite"). LeanExplore's hybrid semantic search, leveraging its AI-generated informal translations and contextual understanding, can retrieve the relevant \code{StatementGroup}. Furthermore, the integrated dependency exploration allows them to click through and understand not only the retrieved statement but also its immediate context, providing a deeper understanding and facilitating faster integration into their ongoing proof.

\subsection{Enhancing learning for new Lean users}
For students or researchers new to Lean and its libraries, grasping complex definitions can be challenging. LeanExplore facilitates understanding through its AI-assisted features. Users can explore declarations and then, via the MCP server \citep{anthropic_model_context_protocol}, integrate LeanExplore with AI assistants like Anthropic's Claude. This allows for intuitive, conversational learning about Lean concepts directly with Claude, which leverages LeanExplore's backend for accurate information retrieval. Such an setup, accessible for instance through Claude's desktop application, provides a dynamic learning experience beyond static documentation and without requiring direct CLI interaction for the chat.

\subsection{Empowering AI agents}
LeanExplore's MCP server is pivotal for AI-driven mathematical research, enabling AI agents to programmatically interact with Lean codebases. For instance, an automated theorem-proving agent can utilize the MCP \code{search} tool to retrieve relevant definitions and theorems from LeanExplore as premises for a given proof goal. This structured data access, combined with planned future enhancements like nightly database updates and broader repository indexing, can empower agents to stay current and even contribute novel formalizations or proofs, potentially via automated pull requests. The efficacy of this approach has been demonstrated in early explorations where Claude Code, an autonomous coding agent by Anthropic, successfully proved theorems by leveraging LeanExplore's MCP interface to query for necessary mathematical facts.

\section{Experimental results}
\label{sec:experimental_results}

To evaluate LeanExplore's practical search efficacy, we conducted a comparative study using a dataset of 300 AI-generated natural language queries (listed in Appendix~\ref{appendix:evaluation_queries}) designed to reflect typical search tasks within the Mathlib library. Each query was executed on three leading search engines: LeanExplore, LeanSearch \citep{Gao2024MathlibSearch}, and Moogle \citep{MoogleAI}. In particular, LeanSearch previously demonstrated itself as being a leading, if not the best, Lean semantic search tool. 

For each query, the top five search results (including the formal Lean code, docstring, and any provided informal statement) from each engine were presented blindly to Google's Gemini 2.5 Flash model, which acted as an automated evaluator. The order of presentation for the three engines' results was permuted for each query to mitigate positional bias. The evaluator LLM was tasked, using the prompt detailed in Appendix~\ref{appendix:llm_prompt}, to rank the sets of results from the three engines, assigning 1st, 2nd, and 3rd place based on which engine it judged provided the most accurate answer to the natural language query. The evaluator was permitted to assign ties if it deemed the results from multiple engines to be of equal quality. This entire ranking procedure was repeated three times for all 300 queries. The results presented in Table~\ref{tab:engine_rankings} and Table~\ref{tab:head_to_head} are the means from these three runs, with uncertainties reported as $\pm$ one standard error (SE). 

Table~\ref{tab:engine_rankings} summarizes the averaged frequency with which each search engine's results were ranked in 1st, 2nd, or 3rd place by the LLM evaluator. Due to the allowance for ties, the sum of percentages for any given rank (e.g., 1st place) across the engines does not necessarily equal 100\%. The data in Table~\ref{tab:engine_rankings} reveals that LeanExplore's results were ranked 1st by the evaluator LLM in an average of 55.4\% of instances, suggesting its hybrid search mechanism was frequently judged to yield the most pertinent results. LeanSearch also demonstrated strong performance, securing 1st place in an average of 46.3\% of cases. Moogle's results were ranked 1st in 12.0\% of instances. Conversely, Moogle was most frequently ranked 3rd (63.2\%), while LeanExplore was least frequently ranked 3rd (10.8\%). It was also noted during evaluation that Moogle's performance may have been impacted by its occasional lack of natural language statements for retrieved declarations, which were sometimes presented as N/A to the evaluator.

Table~\ref{tab:head_to_head} offers a more direct head-to-head comparison, detailing the win and tie rates for each pair of engines, averaged over the three evaluation runs. These findings further quantify the efficacy of LeanExplore's search approach. In direct comparisons, LeanExplore outperformed both LeanSearch and Moogle. This strong performance is notable given that LeanExplore utilizes the BAAI bge-base-en-v1.5 embedding model \citep{BGEManual}, which, at 109 million parameters, is considerably more lightweight than the large E5mistral-7b model used for LeanSearch \citep{Gao2024MathlibSearch}. 

However, this choice of a smaller model presented some trade-offs. LeanExplore occasionally struggled with very general queries, such as "vector space axioms," where LeanSearch performed better. This is because there are no items in the database distinctly labeled as such, and the larger model better retrieved relevant information. LeanExplore's algorithm excelled at retrieving results for more specific user queries.

\begin{table}[h!]
\centering
\caption{Search engine rankings by evaluator LLM (averaged over 3 runs, 900 trials total). Values are mean percentage $\pm$ standard error (SE).}
\label{tab:engine_rankings}
\begin{tabular}{@{}lccc@{}}
\toprule
Engine & 1st place rate (\%) & 2nd place rate (\%) & 3rd place rate (\%) \\
\midrule
LeanExplore & 55.4 $\pm$ 0.7 & 33.8 $\pm$ 0.6 & 10.8 $\pm$ 0.5 \\
LeanSearch  & 46.3 $\pm$ 1.4 & 34.1 $\pm$ 0.7 & 19.6 $\pm$ 1.1 \\
Moogle      & 12.0 $\pm$ 0.6 & 24.8 $\pm$ 0.2 & 63.2 $\pm$ 0.4 \\
\bottomrule
\end{tabular}
\end{table}

\begin{table}[h!]
\centering
\caption{Head-to-head engine comparison (averaged over 3 runs, 900 trials total). Values are mean percentage $\pm$ standard error (SE).}
\label{tab:head_to_head}
\begin{tabular}{@{}llr@{}}
\toprule
Compared engines & Metric & Rate (\%) \\
\midrule
LeanExplore vs LeanSearch & LeanExplore wins & 50.0 $\pm$ 1.8 \\
                         & LeanSearch wins  & 39.4 $\pm$ 1.4 \\
                         & Ties             & 10.6 $\pm$ 0.7 \\
\midrule
LeanExplore vs Moogle    & LeanExplore wins & 79.2 $\pm$ 0.6 \\
                         & Moogle wins      & 15.9 $\pm$ 0.4 \\
                         & Ties             & 4.9 $\pm$ 0.6 \\
\midrule
LeanSearch vs Moogle     & LeanSearch wins  & 72.0 $\pm$ 0.4 \\
                         & Moogle wins      & 23.2 $\pm$ 0.9 \\
                         & Ties             & 4.8 $\pm$ 0.7 \\
\bottomrule
\end{tabular}
\end{table}

\section{Future work and directions}
\label{sec:future_work}
LeanExplore is an actively developing project with several key directions for future enhancement. A primary goal is to ensure the database remains current and comprehensive. We plan to implement nightly updates and significantly broaden the indexed corpus by incorporating a wide array of repositories from platforms like Reservoir \citep{leanfro_reservoir}. This expansion will necessitate a new data extraction methodology capable of handling diverse project versions and resolving inter-project dependencies. Such timely and extensive data access could empower AI agents to rapidly contribute to mathematics by facilitating the generation and submission of daily pull requests with new formalizations or proofs.

Further development will focus on enhancing search capabilities and user choice. We intend to introduce the ability for users to select from different underlying embedding models, catering to varying performance and resource requirements. A feature we are considering is the capacity to search for or by proof states, which would allow users to find existing proofs or lemmas relevant to specific goals, greatly aiding in proof construction. These directions aim to continually improve LeanExplore's utility for both the Lean community and AI-driven mathematical research.

\section{Conclusion}
\label{sec:conclusion}
The rapid expansion and inherent complexity of the Lean 4 ecosystem necessitate sophisticated tools for efficient navigation and knowledge discovery. This paper has introduced LeanExplore, a search engine and toolkit designed to address this challenge. LeanExplore provides robust semantic search across diverse Lean 4 packages, leveraging a hybrid approach that combines multi-source semantic embeddings—derived from formal code, docstrings, AI-generated informal translations, and file path keywords—with lexical BM25+ scores and structural PageRank information. This methodology, centered around the novel \code{StatementGroup} abstraction for user-authored code blocks, allows for nuanced and effective retrieval.

We have detailed LeanExplore's modular architecture, supporting both local data access and a remote API, and its comprehensive data extraction pipeline. This pipeline integrates global project analysis with fine-grained, AST-level information and employs an LLM for inductive informal translation of Lean statements, enriching the basis for semantic understanding. Accessible via a dedicated website, a Python library, and a CLI, LeanExplore offers features like dependency exploration and an interactive AI-assisted chat. Crucially, its model context protocol (MCP) server implementation provides a standardized interface for AI agents, enabling them to programmatically query and utilize the structured knowledge within Lean libraries.

Our experimental results demonstrate LeanExplore's strong performance in retrieving relevant declarations, notably achieving these results with a comparatively lightweight embedding model (109 million parameters). This highlights the efficacy of its overall system design. By offering specialized search, AI-augmented interaction, and robust data processing, LeanExplore has the potential to significantly enhance the productivity of Lean users, lower the entry barrier for newcomers, and, critically, open new avenues for AI research in formal mathematics. As the Lean 4 ecosystem continues to grow, tools like LeanExplore will be increasingly vital for harnessing its full potential and for fostering a new generation of AI-driven mathematical exploration and contribution.

\section*{Acknowledgments} 

The author thanks Eric Taucher and the broader Lean prover community for their valuable feedback on LeanExplore. He is also grateful to Alessandro Selvitella for his extensive mentorship on machine learning methods.

\small

\bibliographystyle{plainnat}
\bibliography{references}

\appendix

\section{Example search result scoring and impact of BM25+}
\label{appendix:scoring_example_comparison}

To illustrate the hybrid ranking mechanism and the specific impact of the BM25+ lexical scoring component \citep{lv2011lower}, this section details the scores and AI-generated informal descriptions for top results retrieved by LeanExplore for the query: \texttt{finite morphism schemes}. 

The weights used for combining the normalized scores are: semantic similarity (1.0) and PageRank (0.2). For the first list presented below, BM25+ also has a weight of (1.0); for the second list, the BM25+ weight is (0.0) to demonstrate its impact. These settings are as described in Section~\ref{sec:search_algorithm}. The raw (unnormalized) score range for the underlying normalized semantic similarity for this specific query was [0.5988, 0.6533]. When active, the raw BM25+ score range was [2.4525, 7.2412]. The raw PageRank score range was [0.0000, 0.4057].

Comparing the rankings with and without the BM25+ component highlights the influence of lexical matching. When BM25+ is active, results with strong keyword relevance to the query "\texttt{finite morphism schemes}" are promoted. For instance, \texttt{AlgebraicGeometry.IsFinite.instIsIntegralHom} ranks higher due to its BM25+ contribution. Conversely, when BM25+ is omitted, items with high semantic similarity but potentially weaker keyword matches for this specific query, such as \texttt{AlgebraicGeometry.Scheme.Hom.toRationalMap}, rise in the rankings. This demonstrates that while semantic similarity and PageRank provide a strong relevance baseline, BM25+ refines the order by incorporating lexical term importance, leading to the hybrid system's final ranking. The top result from the full hybrid search, \texttt{AlgebraicGeometry.IsFinite.instLocallyOfFiniteType}, maintains a high rank in both scenarios due to its strong performance across multiple signals.

\subsection*{Results with hybrid scoring}

Below are the top 5 results when all three scoring components (semantic similarity, BM25+, PageRank) are active, ranked by their total weighted score.

\begin{enumerate}
    \item \textbf{Name:} \texttt{AlgebraicGeometry.IsFinite.instLocallyOfFiniteType}
    \begin{itemize}
        \item \textbf{Informal description:} A finite morphism of schemes is locally of finite type.
        \item \textbf{Weighted semantic score:} 1.0000
        \item \textbf{Weighted BM25+ score:} 0.9622
        \item \textbf{Weighted PageRank score:} 0.0050
        \item \textbf{Total weighted score:} 1.9672
    \end{itemize}
    \vspace{0.5em}

    \item \textbf{Name:} \texttt{AlgebraicGeometry.IsFinite.instIsIntegralHom}
    \begin{itemize}
        \item \textbf{Informal description:} A finite morphism of schemes is an integral morphism.
        \item \textbf{Weighted semantic score:} 0.8840
        \item \textbf{Weighted BM25+ score:} 0.9415
        \item \textbf{Weighted PageRank score:} 0.0035
        \item \textbf{Total weighted score:} 1.8290
    \end{itemize}
    \vspace{0.5em}

    \item \textbf{Name:} \texttt{AlgebraicGeometry.instIsProperOfIsFinite}
    \begin{itemize}
        \item \textbf{Informal description:} A finite morphism of schemes is proper.
        \item \textbf{Weighted semantic score:} 0.8202
        \item \textbf{Weighted BM25+ score:} 0.9007
        \item \textbf{Weighted PageRank score:} 0.0000
        \item \textbf{Total weighted score:} 1.7209
    \end{itemize}
    \vspace{0.5em}
    
    \item \textbf{Name:} \texttt{AlgebraicGeometry.IsProper.instOfIsFinite}
    \begin{itemize}
        \item \textbf{Informal description:} A finite morphism of schemes is proper.
        \item \textbf{Weighted semantic score:} 0.8202
        \item \textbf{Weighted BM25+ score:} 0.8683
        \item \textbf{Weighted PageRank score:} 0.0028
        \item \textbf{Total weighted score:} 1.6913
    \end{itemize}
    \vspace{0.5em}

    \item \textbf{Name:} \texttt{AlgebraicGeometry.IsFinite.iff\_isIntegralHom\_and\_locallyOfFiniteType}
    \begin{itemize}
        \item \textbf{Informal description:} A morphism of schemes is finite if and only if it is an integral morphism and locally of finite type.
        \item \textbf{Weighted semantic score:} 0.7204
        \item \textbf{Weighted BM25+ score:} 0.9275
        \item \textbf{Weighted PageRank score:} 0.0083
        \item \textbf{Total weighted score:} 1.6562
    \end{itemize}
    \vspace{0.5em}
\end{enumerate}

\subsection*{Results with scoring without BM25+}

Below are the top 5 results for the same query when the BM25+ component's weight is set to 0. The ranking is now determined by the sum of weighted semantic similarity and weighted PageRank. Note the change in ranking and total scores compared to the list above.

\begin{enumerate}
    \item \textbf{Name:} \texttt{AlgebraicGeometry.IsFinite.instLocallyOfFiniteType}
    \begin{itemize}
        \item \textbf{Informal description:} A finite morphism of schemes is locally of finite type.
        \item \textbf{Weighted semantic score:} 1.0000
        \item \textbf{Weighted PageRank score:} 0.0050
        \item \textbf{Total weighted score:} 1.0050
    \end{itemize}
    \vspace{0.5em}

    \item \textbf{Name:} \texttt{AlgebraicGeometry.Scheme.Hom.toRationalMap}
    \begin{itemize}
        \item \textbf{Informal description:} A morphism of schemes from $X$ to $Y$ can be viewed as a rational map from $X$ to $Y$ by first considering the morphism as a partial map defined on all of $X$, and then viewing this partial map as a rational map.
        \item \textbf{Weighted semantic score:} 0.9188
        \item \textbf{Weighted PageRank score:} 0.0026
        \item \textbf{Total weighted score:} 0.9214
    \end{itemize}
    \vspace{0.5em}

    \item \textbf{Name:} \texttt{AlgebraicGeometry.Scheme.Hom.toPartialMap}
    \begin{itemize}
        \item \textbf{Informal description:} Given a morphism $f$ of schemes, we can view it as a partial map from $X$ to $Y$ whose domain is the entire scheme $X$. This partial map is represented by the entire scheme $X$ (as the domain), the fact that $X$ is a dense subset of itself, and the morphism itself, precomposed with the canonical isomorphism from $X$ to the restriction of $X$ to the open subset $X$. In other words, the partial map is just $f$.
        \item \textbf{Weighted semantic score:} 0.8860
        \item \textbf{Weighted PageRank score:} 0.0099
        \item \textbf{Total weighted score:} 0.8959
    \end{itemize}
    \vspace{0.5em}
    
    \item \textbf{Name:} \texttt{AlgebraicGeometry.IsFinite.instIsIntegralHom}
    \begin{itemize}
        \item \textbf{Informal description:} A finite morphism of schemes is an integral morphism.
        \item \textbf{Weighted semantic score:} 0.8840
        \item \textbf{Weighted PageRank score:} 0.0035
        \item \textbf{Total weighted score:} 0.8875
    \end{itemize}
    \vspace{0.5em}

    \item \textbf{Name:} \texttt{AlgebraicGeometry.Scheme.Cover.Hom.comp}
    \begin{itemize}
        \item \textbf{Informal description:} Given two composable morphisms $f$ and $g$ between covers of a scheme with respect to a morphism property $P$ that is stable under composition, their composition is the morphism defined by mapping an index $j$ in $\mathcal{U}$ to the index $g(f(j))$ in $\mathcal{W}$ and by mapping an open subscheme $\mathcal{U}_j$ to $\mathcal{W}_{g(f(j))}$ via the composition of $\mathcal{U}_j \to \mathcal{V}_{f(j)}$ and $\mathcal{V}_{f(j)} \to \mathcal{W}_{g(f(j))}$.
        \item \textbf{Weighted semantic score:} 0.8191
        \item \textbf{Weighted PageRank score:} 0.0042
        \item \textbf{Total weighted score:} 0.8233
    \end{itemize}
    \vspace{0.5em}
\end{enumerate}

\section{LLM evaluation prompt}
\label{appendix:llm_prompt}

The following prompt was used to instruct Google's Gemini 2.5 Flash model to evaluate and rank the search results from LeanExplore, LeanSearch, and Moogle. The placeholders \code{Engine A}, \code{Engine B}, and \code{Engine C} were randomly assigned to the three search engines for each query to mitigate bias. The \code{\{Query\}} and \code{\{Formatted Results for Engine X\}} placeholders were populated with the specific query and the top 5 formatted results from the respective engine.

\begin{lstlisting}[caption={LLM Evaluation Prompt}, label={lst:llm_eval_prompt}, language={}, basicstyle=\ttfamily\scriptsize, breaklines=true, breakatwhitespace=true, xleftmargin=2em, frame=single, numbers=left, stepnumber=1, numberstyle=\tiny\color{gray}]
System Prompt:
You are an expert search result evaluator. Your task is to analyze search results from three different search engines for a given query. These engines are presented as Engine A, Engine B, and Engine C. You need to rank these three engines from best to worst based on how accurate the search results are for the provided query. If two or more engines are of comparable quality for a given rank, you can declare them as tied.

First, provide your detailed reasoning for the ranking. After your reasoning, on a new and final line, provide your ranking. Use the placeholders 'Engine A', 'Engine B', 'Engine C'. Separate distinct ranks with a comma and a space (e.g., 'Engine A, Engine B'). Separate tied engines at the same rank with an equals sign and spaces around it (e.g., 'Engine A = Engine B'). All three engine placeholders must be present in the ranking line.
Examples of valid ranking lines:
- Engine B, Engine A, Engine C
- Engine A = Engine B, Engine C
- Engine A, Engine B = Engine C
- Engine A = Engine B = Engine C

User Prompt:
Original Query: "{Query}"

Search Results:

----- Engine A -----

{Formatted Results for Engine A:
RESULT 1
Code:
{Code for Result 1, Engine A}
Docstring:
{Docstring for Result 1, Engine A}
Informal description:
{Informal Description for Result 1, Engine A}

RESULT 2
Code:
{Code for Result 2, Engine A}
Docstring:
{Docstring for Result 2, Engine A}
Informal description:
{Informal Description for Result 2, Engine A}
... (up to 5 results)
}

----- Engine B -----

{Formatted Results for Engine B:
RESULT 1
Code:
{Code for Result 1, Engine B}
Docstring:
{Docstring for Result 1, Engine B}
Informal description:
{Informal Description for Result 1, Engine B}
... (up to 5 results)
}

----- Engine C -----

{Formatted Results for Engine C:
RESULT 1
Code:
{Code for Result 1, Engine C}
Docstring:
{Docstring for Result 1, Engine C}
Informal description:
{Informal Description for Result 1, Engine C}
... (up to 5 results)
}

Based on the original query, please provide your detailed reasoning for ranking Engine A, Engine B, and Engine C. After your reasoning, on a new and final line, state your ranking of Engine A, Engine B, and Engine C from best to worst. Use only the placeholders 'Engine A', 'Engine B', 'Engine C'.
If engines are tied, list them separated by ' = ' (e.g., 'Engine A = Engine B').
If ranks are distinct, separate them by ', ' (e.g., 'Engine A, Engine B').
This final line must contain only the ranking and all three engine placeholders.
For example:
Engine B, Engine A, Engine C
OR
Engine A = Engine B, Engine C
OR
Engine A, Engine B = Engine C
OR
Engine A = Engine B = Engine C
\end{lstlisting}

The formatting for each result within \code{\{Formatted Results for Engine X\}} included the Lean code (or statement text), its docstring (if available, with Moogle's docstrings being pre-processed to remove comment markers), and an informal natural language description (if available). If any of these components were not available for a specific result, "N/A" was used.

\section{List of evaluation queries}
\label{appendix:evaluation_queries}

The following queries were used in the evaluation described in Section \ref{sec:experimental_results}:

group definition, ring definition, natural numbers type, addition commutativity theorem, topological space definition, first isomorphism theorem groups, vector space axioms, category definition, fundamental theorem of calculus, prime number definition, finite sets properties, Schroeder-Bernstein theorem, manifold definition, fiber bundle, limit definition analysis, continuous functions theorems, Heine-Borel theorem, measurable space definition, sigma-algebra definition, dominated convergence theorem, polynomial definition, binomial theorem, scheme algebraic geometry, module over ring, Chinese Remainder Theorem, function derivative definition, chain rule differentiation, integral definition, mean value theorem, metric space definition, Cauchy sequence, real numbers completeness, Baire category theorem, Hilbert space definition, Riesz representation theorem, Banach space definition, Hahn-Banach theorem, open mapping theorem Banach spaces, closed graph theorem, spectral theorem self-adjoint operators, Lie group definition, Lie algebra definition, exponential map Lie theory, universal enveloping algebra, Poincare-Birkhoff-Witt theorem, homological complex definition, homology, snake lemma, long exact sequence homology, functor definition, natural transformations, Yoneda lemma, adjoint functors, monad definition, continuous functions properties, intermediate value theorem, Lagrange's theorem groups, ideals properties, vector space dimension, eigenvalues eigenvectors, compact sets properties, tangent spaces, inverse function theorem, implicit function theorem, Stokes' theorem, Cauchy's integral theorem, residue theorem, Turing machine, halting problem, P vs NP, probability space definition, law of large numbers, central limit theorem, Bayes' theorem, L'Hopital's Rule, Taylor series expansion, Fourier transform, Hilbert space properties, Lie algebra of Lie group, Peano axioms, Zorn's lemma, Well-ordering principle, Cantor's theorem uncountability reals, Brouwer fixed point theorem, Banach fixed point theorem, field definition, Galois theory, fundamental theorem of algebra, tensor product modules, exact sequences, homology cohomology, natural numbers definition, induction principle, pigeonhole principle, Euclidean algorithm, Fermat's Little Theorem, Euler's totient theorem, quadratic reciprocity, complex numbers definition, De Moivre's formula, Cauchy-Riemann equations, measure definition, Lebesgue integration, Fubini's theorem, Radon-Nikodym theorem, graph definition, Ramsey theory, partial order definition, lattices Boolean algebras, Stone's representation theorem Boolean algebras, filter definition, ultrafilter lemma, Tychonoff's theorem, Sard's theorem, Whitney embedding theorem, simplicial complex definition, Sylow theorems, free group, Noetherian ring, localization of module, prime ideal maximal ideal, unique factorization domain, Eisenstein's criterion, Hilbert's basis theorem, tensor algebra vector space, exterior algebra, Clifford algebra, algebraic closure, Galois correspondence, splitting field, derived functors homological algebra, sum of four squares theorem, Pell's equation, Bernoulli numbers zeta function, Hensel's lemma, Liouville's theorem Diophantine approximation, finiteness ideal class group, Dirichlet's unit theorem, Schur's lemma representation theory, basis free module, dual space infinite dimensional vector space, structure theorem finitely generated modules PID, Cayley-Hamilton theorem, Gram-Schmidt process, diagonalization symmetric matrices, Stone-Cech compactification, Urysohn's lemma normal spaces, Stone-Weierstrass theorem, uniform convergence functions, completion metric space, topological group, Haar measure, continuous linear maps topological vector spaces, locally convex topological vector spaces, Banach-Steinhaus theorem, orthogonal projection, Lax-Milgram theorem, smooth function C-infinity, Jensen's inequality integrals, Caratheodory's theorem convex hull, Borel sets sigma-algebras, Bochner integral, change of variables formula multiple integrals, divergence theorem, Liouville's theorem entire functions, maximum modulus principle, Schwarz lemma, removable singularities complex analysis, Schwartz space, tempered distributions Fourier transform, Fourier inversion formula, Riemann-Lebesgue lemma, Parseval's theorem Fourier series, Kolmogorov's zero-one law, moment generating function moments, Markov inequality, Chebyshev's inequality, strong law of large numbers, martingale convergence theorem, optional stopping theorem, hitting time stochastic processes, Lie bracket vector fields, flow vector field, pullback vector bundle, Zariski topology affine space, Nullstellensatz, Hall's marriage theorem, Catalan numbers, Bell numbers, VC dimension, Turan's theorem, van der Waerden's theorem, Roth's theorem arithmetic progressions, Cauchy-Davenport theorem, omega-limit sets dynamical systems, Ackermann function, Rice's theorem, pumping lemma context-free languages, ordinal numbers, cardinal arithmetic, model theory compactness theorem, Lowenheim-Skolem theorem, comma category, limits colimits, sheafification, monoidal category, cartesian closed category, abelian category homological algebra, p-adic integers, hyperreal numbers, abelianization group, structure theorem finitely generated abelian groups, first isomorphism theorem rings, polynomial root multiplicity, perfect closure field characteristic p, functorial homology, Dirichlet class number formula, orthogonality characters group representations, noetherian module, alternating map, isomorphism finite dimensional space bidual, minimal polynomial linear operator, polar form quadratic form, Heine-Cantor theorem uniform continuity, topological ring, topological module, equivalence norms finite dimensions, isometric inclusion double dual normed space, Hilbert basis orthonormal basis, Leibniz formula derivatives, local extrema, product measures, convolution functions, independence random variables, Bernoulli law, integral curves vector fields, locally ringed space schemes, adjacency matrix graph, Hales-Jewett theorem, Ruzsa triangle inequality, fixed points dynamical system, polynomial-time computation, model ZFC set theory, elementary embedding model theory, equivalence of categories, representable functor, Dedekind cuts real numbers, solvable group, derived series group theory, group cohomology, Jacobson radical ring, Dedekind domain, completion local ring, Gauss's lemma polynomial irreducibility, symmetric polynomial theorem, Galois field GF(p\textasciicircum n), primitive element theorem field extensions, exact functor, projective resolution module, injective resolution, Tor functor, Ext functor, Dedekind zeta function, local field, global field, injective module, flat module, Jordan canonical form, singular value decomposition, metrization theorems topology, compact-open topology function spaces, reflexive Banach space, distributional derivative, harmonic function, Riemann mapping theorem, Picard's great theorem essential singularities, Borel-Cantelli lemma, characteristic function random variable, Levy continuity theorem characteristic functions, Brownian motion, Poisson process, Riemannian manifold, geodesic equation, Riemann curvature tensor, differential forms exterior derivative, de Rham cohomology, generating functions combinatorial sequences, max-flow min-cut theorem, ergodic theorem, invariant measure dynamical system, Poincare recurrence theorem, lambda calculus reduction rules, Church-Rosser property, intuitionistic logic classical logic, sheaf cohomology, Brauer group field, central simple algebra, Grothendieck topology sheaves site, Ito calculus stochastic differential equations, Krull dimension.

\end{document}